\documentclass[letterpaper, 10 pt, conference]{ieee/ieeeconf}  

\IEEEoverridecommandlockouts                              
\usepackage{graphics} 
\usepackage{epsfig} 
\usepackage{mathptmx} 
\usepackage{times} 
\usepackage{amsmath} 
\usepackage{amssymb}  
\usepackage[usenames, dvipsnames]{color}
\usepackage[linesnumbered,ruled,vlined]{algorithm2e}
\usepackage[caption=false,font=footnotesize]{subfig}
\usepackage[export]{adjustbox}
\usepackage{graphicx}
\usepackage{float}
\usepackage{multirow}

\definecolor{purple}{RGB}{149,33,246}

\newcommand{\YOON}[1] {
	\textcolor{red}{\bfseries{YOON: {#1}}}
}

\newcommand{\IK}[1] {
	\textcolor{blue}{\bfseries{IK: {#1}}}
}

\newcommand{\MB}[1] {
	\textcolor{purple}{\bfseries{MB: {#1}}}
}

\newcommand{\Dinesh}[1] {
	\textcolor{Green}{\bfseries{Dinesh: {#1}}}
}

\renewcommand{\baselinestretch}{1}

\providecommand{\shortcite}[1]{\cite{#1}}
\DeclareMathOperator*{\argmin}{arg\,min}

\title{\LARGE \bf
Reflection-Aware Sound Source Localization}

\newcommand{\Skip}[1]{}
\renewcommand{\paragraph}[1]{{\bf {#1}}} 

\begin{document}

\author{Inkyu An$^{1}$, Myungbae Son$^{2}$, Dinesh Manocha$^3$, and Sung-eui Yoon$^{4}$ \\
{\small (Video is available at https://youtu.be/TkQ36lMEC-M)}
\thanks{$^{1}$Inkyu An, 
$^{2}$Myungbae Son, and $^{4}$Sung-eui Yoon are at KAIST, South Korea;
$^3$Dinesh Manocha is at UNC-Chapel Hill, USA;
       {\tt\small inkyu\_ahn, nedsociety@kaist.ac.kr, dmanocha@gmail.com, sungeui@kaist.edu}}%
}

\maketitle
\thispagestyle{empty}
\pagestyle{empty}

\begin{abstract}
We present a novel, reflection-aware method for 3D sound localization in indoor environments.  Unlike prior approaches, which are mainly based on continuous sound signals from a stationary source, our formulation is designed to localize the position instantaneously from signals within a single frame. We consider direct sound and indirect sound signals that reach the microphones after reflecting off surfaces such as ceilings or walls. We then generate and trace direct and reflected acoustic paths using inverse acoustic ray tracing and utilize these paths with Monte Carlo localization to estimate a 3D sound source position. We have implemented our method on a robot with a cube-shaped microphone array and tested it against different settings with continuous and intermittent sound signals with a stationary or a mobile source. Across different settings, our approach can localize the sound with an average distance error of 0.8~m tested in a room of 7~m by 7~m area with 3~m height, including a mobile and non-line-of-sight sound source. We also reveal that the modeling of indirect rays increases the localization accuracy by 40\% compared to only using direct acoustic rays.
\end{abstract}

\section{INTRODUCTION}
\label{sec:1}

Robots are increasingly used in our daily environments, and the demands on robots to
interact with humans and the environment using acoustic cues are getting
stronger. The recent popularity of intelligent devices such as  Amazon Echo and Google
Home is giving rise to new challenges in acoustic scene analysis.  One of the
key issues in these applications  is localizing the exact position of a sound
source in the real world.  Once a robot identifies the location of the sound
source, it can approach the location and perform many useful tasks.  The
resulting problem, \emph{sound source localization} (SSL), has been well-formulated
and well-studied for decades~\cite{brandstein2013microphone}.

Most prior work in SSL has been related to the design of microphone arrays and the use of digital signal processing techniques.  Nonetheless, it remains a challenging problem to exactly locate the sound source with limited information available from the sensors equipped on a robot. In the most general setting,  the localization problem tends to be ill-posed. Most of the research in the last two decades has been dedicated to capturing the local characteristics of input signals, such as incoming directions of a sound. Specifically, Time Difference of Arrival (TDOA) based SSL techniques have been investigated for the last two decades, and mainly utilize the difference of arrival time between two microphone pairs~\cite{knapp1976generalized,valin2007robust}. In most cases, they are successfully used to detect the direction of the incoming sound signal, but not the position of the sound source that generated those signals.

\Skip{
We explain an example why it is important to know the exact position of a sound
source in the real world.  Assuming that there are a robot and a person in a
house, the person wants to ask a robot to bring me water.  Without using sound,
the person has to go near the robot and then give an order or send information
using mobile devices like a cell phone.  On the other hand, with using sound,
the person can easily ask the robot to bring me water without any other
unnecessary conditions.  Therefore, 

Nonetheless, SSL has been remaining a challenging problem,
mainly due to its nature of ill-posedness.

}

Recent studies in SSL methods have advanced into addressing the localization
issues under certain
configurations~\cite{sasaki2016probabilistic,su2017towards}. Unfortunately,
their methods require accumulating the incoming sensor data measured from
different locations and orientations.
As a result, these techniques typically assume that a stationary sound source
generates continuous sound signals and that there are no obstacles between the
source and the receiver.

\Skip{
These methods could localize a sound source in specific configurations, where
the data collection hardware has to move to accumulate  the incoming 
sound signals from a stationary source emitting continuous sound
signals. They assume that the convergence of accumulated multiple direction
does not change [cITATTIONs FOR THIS WORK].  YOU ONLY MAKE A CASE FOR HANDLING
NON-CONTINUOUS SIGNALS; BUT YOU DO NOT MOTIVATE THE USE OF SOUND REFLECTIONS OR
HANDLING OF INDIRECT SOUND SIGNALS?
and a sound source has to remain static with continuous sound signal because the convergence of accumulated multiple directions must not be changed.
}

\begin{figure}[t]
	\centering
	\includegraphics[width=0.9\columnwidth]{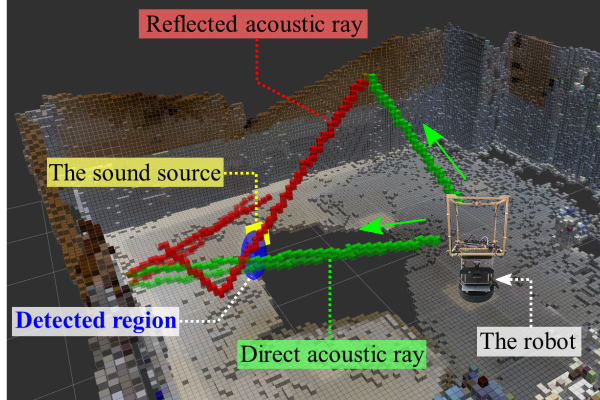}
	\vspace*{-0.3cm}
	\caption{
		Our robot, equipped with a cube-shaped microphone array, localizes a sound source position in the 3D space.  Our formulation takes both direct and indirect sounds into account. Direct acoustic rays (shown in green) are propagated using backward ray tracing based on received signals using a TDOA-based method. Reflected (or indirect) rays (shown in red) are then generated once they hit the boundaries of the reconstructed scene. The blue disk, which is very close to the ground truth,  represents a 95\% confidence ellipse for the estimated sound source, computed by our method. The use of reflected rays improves the localization accuracy by 40\% over only using direct rays.
	}
	\label{fig:resultOfAcousticRay}
	\vspace{-1em}
\end{figure}

\begin{figure*}[t]
	\centering
	\includegraphics[width=2\columnwidth]{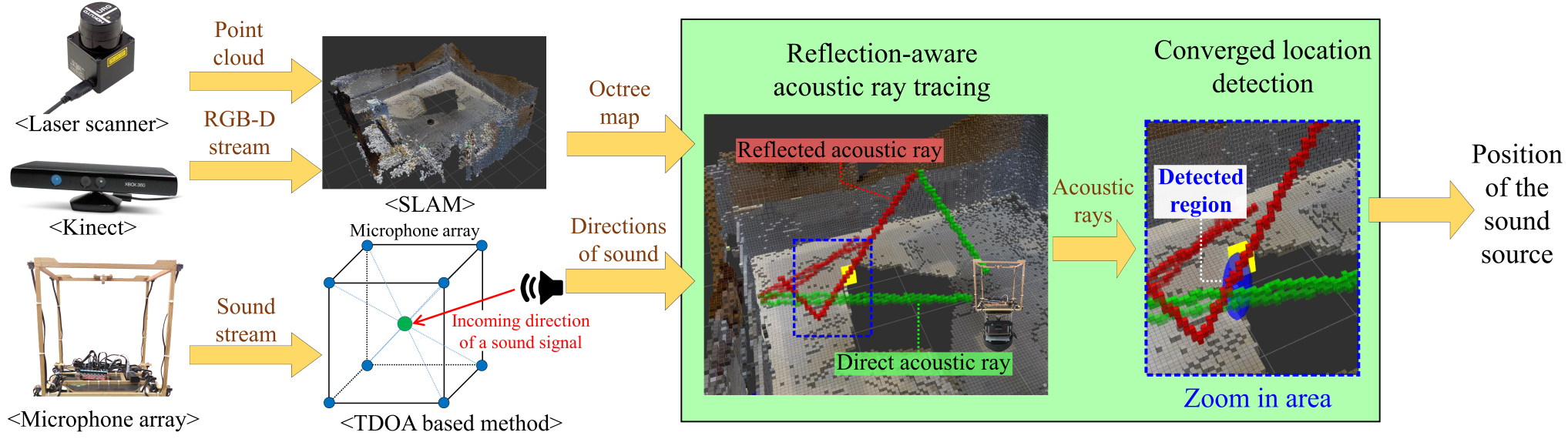}
	\vspace*{-0.3cm}	
	\caption{
		This figure shows an overview of our reflection-aware sound
		source localization approach. Two highlighted modules are our
		main contributions.
	}
	\vspace{-1em}
	\label{fig:blockDiagram}
\end{figure*}

\paragraph{Main contributions.}
We present a novel, reflection-aware SSL approach to localize a 3D position of a sound source in indoor environments.  A key aspect of our work is to model the propagation of sound in the environment. We consider both direct signals between the source and the receiver and indirect signals, which are generated by reflections from the environment such as the wall and ceilings. Specifically, we reconstruct the environment in a voxel-based octree and perform acoustic ray tracing, where direct acoustic rays are generated from signals collected using the TDOA-based method (Sec.~\ref{sec:4_a}). Our acoustic ray tracing models higher orders of reflection, simulating interactions with the boundaries of the environment. We then localize the source by generating hypothetical estimates on these acoustic paths using Monte Carlo localization (Sec.~\ref{sec:4_b}).  

Our approach for modeling the reflections is near real-time and can also handle moving sources as well as non-line-of-sight sources.  Furthermore, our approach can handle intermittent sound signals in addition to continuous ones. We evaluate the performance on three different benchmarks in a classroom environment and test our method with a cube-shaped microphone array mounted on a mobile robot. Given the test environment of 7~m by 7~m area with 3~m height, our method achieves a low average error, e.g., 80~cm, even with a moving sound source and an obstacle occluding the line-of-sight between the listener and the source. This accuracy is achieved by considering higher order reflections in addition to the direct rays. 

\section{RELATED WORK}
\label{sec:2}

In this section, we discuss prior work on sound source localization methods and
sound propagation techniques.

\paragraph{Sound source localization.}
There is considerable work on localizing the sound source using a microphone array~\cite{brandstein2013microphone}.  The vast majority of existing sound source localization (SSL) methods focus on accurately detecting only the incoming directions of the sound. Many methods are based on  TDOA between two microphone pairs. Generalized cross-correlation with phase transform~\cite{knapp1976generalized} is a well-known method for performing  TDOA estimation. Nakamura et al.~\shortcite{nakamura2009intelligent} overcome the noise weakness in dynamic environments by selecting specific sound signals to cancel or focus. Valin et al.~\shortcite{valin2007robust} use a beam-forming technique to perform robust sound source localization. Other methods use multiple signal classification techniques to isolate the number of sound sources~\cite{schmidt1986multiple,argentieri2007broadband,otsuka2011bayesian}.

TDOA techniques are capable of classifying the incoming directions of the prominent sound signals.  Recent efforts have been directed at overcoming this limitation and locating the sound source exactly. Ishi et al.~\shortcite{ishi2013using} present a method for estimating 3D sound source locations by integrating the sound directions measured from multiple microphone arrays, which are installed in fixed positions of a room. Narang et al.~\shortcite{narang2014auditory} suggest a 2D reflection-robust SSL method using visual simultaneous localization and mapping (SLAM).  They gather the sound vectors per frame on a visual odometry made by visual SLAM and try to find an intersection point between them. In recent work, Sasaki et al.~\shortcite{sasaki2016probabilistic} devise a 3D sound source discovery system from a moving microphone array. As they move around with a hand-held unit, they compute the planes that contain the direction of the sound and choose the convergence region among the planes using the particle filter.

In general, computing the exact location of the sound source is inherently an ill-posed problem~\cite{brandstein2013microphone}, and thus most of these prior work operates under some common assumptions about the sound patterns or signals. Notably, the sound sources are assumed to be persistent and stationary, which allows the accumulation of temporal data over time using mobile microphones. Our method, however, is designed to be more general; it requires much less information captured from a single frame and can handle a moving sound source without a line-of-sight from the listener.

\paragraph{Sound propagation.}
Various methods have been proposed to simulate the propagation of sounds. A recent survey is given in~\cite{Savioja15} and many issues in their application to real-world scenes are addressed in~\cite{Vorlander13}. At a broad level, sound propagation techniques are categorized as {Numerical Acoustics} (NA) and {Geometric Acoustics (GA)} techniques. NA methods try to simulate an exact acoustic wave equation and compute an accurate solution. However, the complexity of these algorithms can increase as a fourth power of the maximum frequency of the simulation. In practice, they are limited to low-frequency sources and offline computations. 
On the other hand, the GA methods are based on ray tracing and its variants. They assume that sound waves travel in straight lines and bounce off the boundaries~\cite{kuttruff2007acoustics}. This approximation is valid for high-frequency sounds, but these methods are unable to accurately model low-frequency effects like diffraction. There is extensive work on developing interactive sound simulation algorithms based on ray tracing that can also handle dynamic environments~\cite{vorlander1989,Schissler16}.  Our inverse acoustic ray tracing method is developed based on these algorithms.

\section{Overview}
\label{sec:3}

In this section, we explain the context for our problem and give an overview of our approach.  Sound source location (SSL) has been studied and most prior methods for acoustic scene analysis are mainly used to identify the incoming sound directions. Since the most general version of SSL is an ill-posed problem, we narrow down our scope by making some assumptions about the source and the indoor environment. 

In this work, we focus on localizing a sound source for real-time applications
and mainly consider direct and reflected sound signals in 3D scenes that are
captured using a microphone array.  We assume that original sound signals from
a sound source are high-frequency sound waves (e.g., clapping sound) so that
our ray tracing based model is accurate.
In a similar spirit, we focus on indoor environments, where the walls and
ceilings consist of diffuse and specular acoustic materials. In our current
approach, we mainly model the specular reflections that carry relatively high
energy.

Given such an environment, we present a novel reflection-aware SSL algorithm
for accurately localizing a 3D position of a sound source.  At a high level,
our method uses two main components. Given incoming sound signals, we perform
inverse acoustic ray tracing for tracking direct and reflected sound paths.
Next, we identify a 3D location of the sound source by computing a convergence
point of those traced paths in the 3D space
(Fig.~\ref{fig:resultOfAcousticRay}).

Our overall approach is shown in Fig.~\ref{fig:blockDiagram}.  The input sound
signals are collected via multiple (e.g., eight) microphones in a microphone
array and evaluated using a TDOA (Time Difference Of Arrival) based method. The
TDOA algorithm evaluates the input sound directions, along with their
intensities and representative frequencies. Since these sound directions are
not yet classified as corresponding to direct or reflected directions of sound
paths, we use acoustic ray tracing to evaluate their characteristics.

To obtain the necessary information required to perform acoustic ray
tracing, we also utilize a SLAM module and an octree-based occupancy map to
compute and represent a reconstructed 3D environment and compute the current
position of the robot.

\section{Reflection-Aware SSL}
\label{sec:4}

\begin{table}[t]
	\caption{
		This table lists commonly appearing notations.
	}
	\centering
	\renewcommand{\arraystretch}{1.1}
	\begin{tabular}{p{1.5cm}|p{6.2cm}}
		\hline
		\textbf{Symbol} & \textbf{Description} \\
		\hline
		\hline
		${\dot o}_m$ & The position of the microphone array.\tabularnewline
		\hline
		$(\hat{v}_n, f_n, i_n^k)$ & An incoming direction, frequency and initial energy\tabularnewline
		&  of the $n$-th sound signal, respectively.\tabularnewline
		\hline
		$N$ & The number of sound signals at current time frame.\tabularnewline
		\hline
		$R_n$, $r_n^k$, $\hat{d}_n$ & A ray path traced from $n$-th sound
		signal, and its $k$-th order reflected ray with its directional unit vector.\tabularnewline
		\hline
		$I_n^k(l^\prime)$ & An energy of the sound ray $r_n^k$ at $l = l^\prime$.\tabularnewline
		\hline
		$\alpha(f_n), \alpha_s(f_n)$ & Attenuation coeff. of the
		air, and absorption coeff. of the reflection. \tabularnewline
		\hline
		${\dot p}_{hit}, P_{local}$ & A voxel that is hit by a ray, and
		its local, occupied voxels.\tabularnewline
		\hline
		$\hat{n}$ & A normal vector of a surface locally fit at ${\dot
			p}_{hit}$.\tabularnewline
		\hline
		$\chi_t, x_t^i$ & A set of $W$ particles, and its $i$-th particle at iteration $t$. \tabularnewline
		\hline
	\end{tabular}\quad
	\renewcommand{\arraystretch}{1}
	\label{tbl:notations}
	\vspace{-0.5cm}
\end{table}

In this section, we first explain our acoustic ray tracing, which generates and
traces acoustic paths, while handling reflections.
We then explain how to localize a sound source given those generated acoustic
paths.
Notations used in the rest of the paper are summarized in Table~\ref{tbl:notations}.

\subsection{Acoustic Ray Tracing}
\label{sec:4_a}

We now explain the process of constructing the ray path over the reconstructed scene.  As shown in the overview of our algorithm (Fig.~\ref{fig:blockDiagram}), we first utilize a TDOA based SSL approach for computing incoming sound directions.  These sound signals heard from the detected directions may come directly from the sound source or be reflected from obstacles. While we cannot discern their types exactly at this point, we utilize these incoming directions by generating acoustic rays along these directions, finding useful information about where the sound source is located.

The main observation for our reflection-aware SSL is that, when we generate acoustic rays in reverse directions of the incoming sound, those rays can be propagated and reflected by some objects in the 3D space. Furthermore, when those rays are coming from the same sound source, they converge in a particular location in the 3D space, which is highly likely to be the original sound source location.

\begin{figure*}[t]
	\centering
	\subfloat[Initializing an acoustic ray]{\includegraphics[width=1.5in]{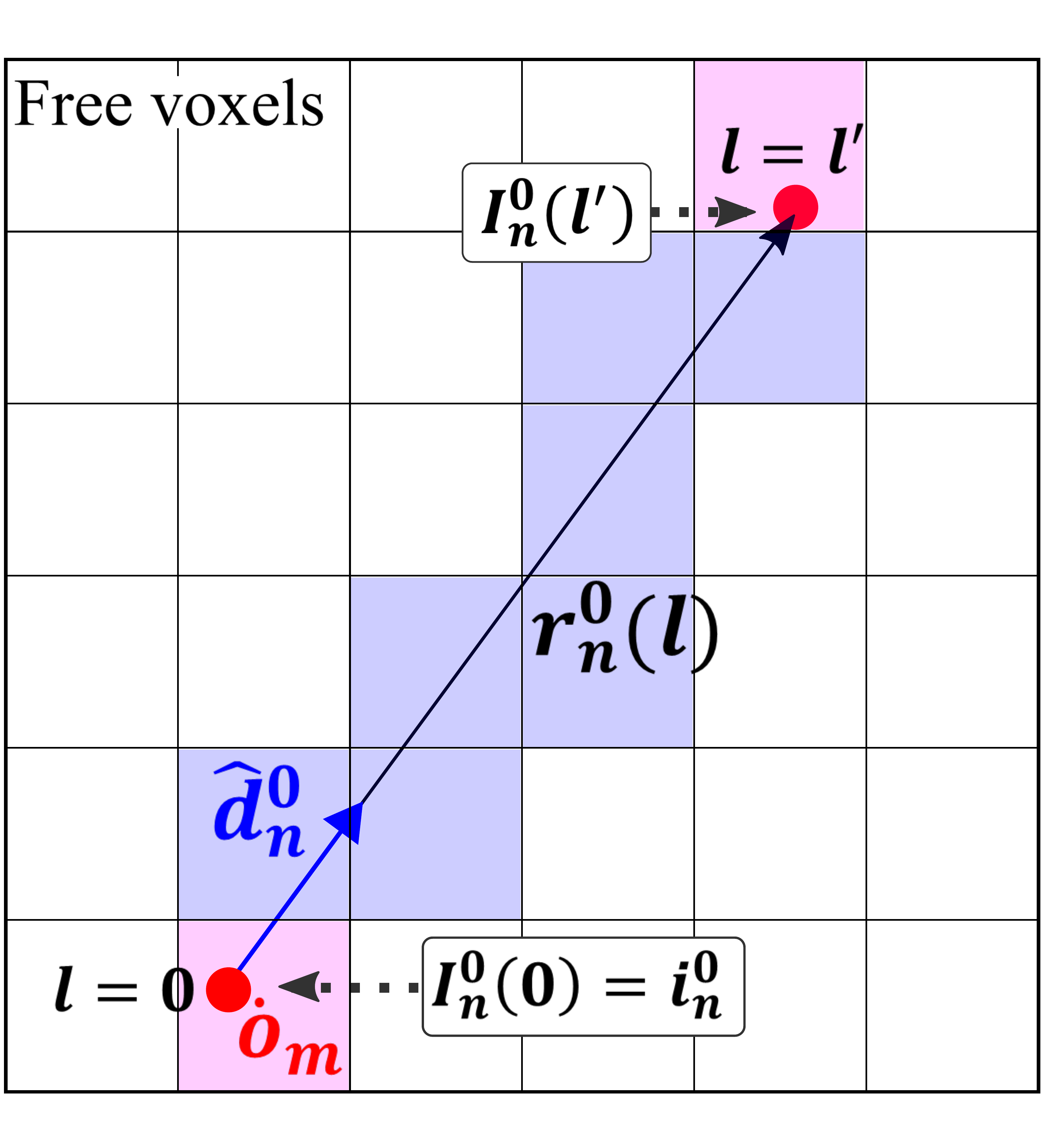}\label{fig:acousticRayTracing_init}}\qquad
	\subfloat[Detecting a hit]{\includegraphics[width=1.5in]{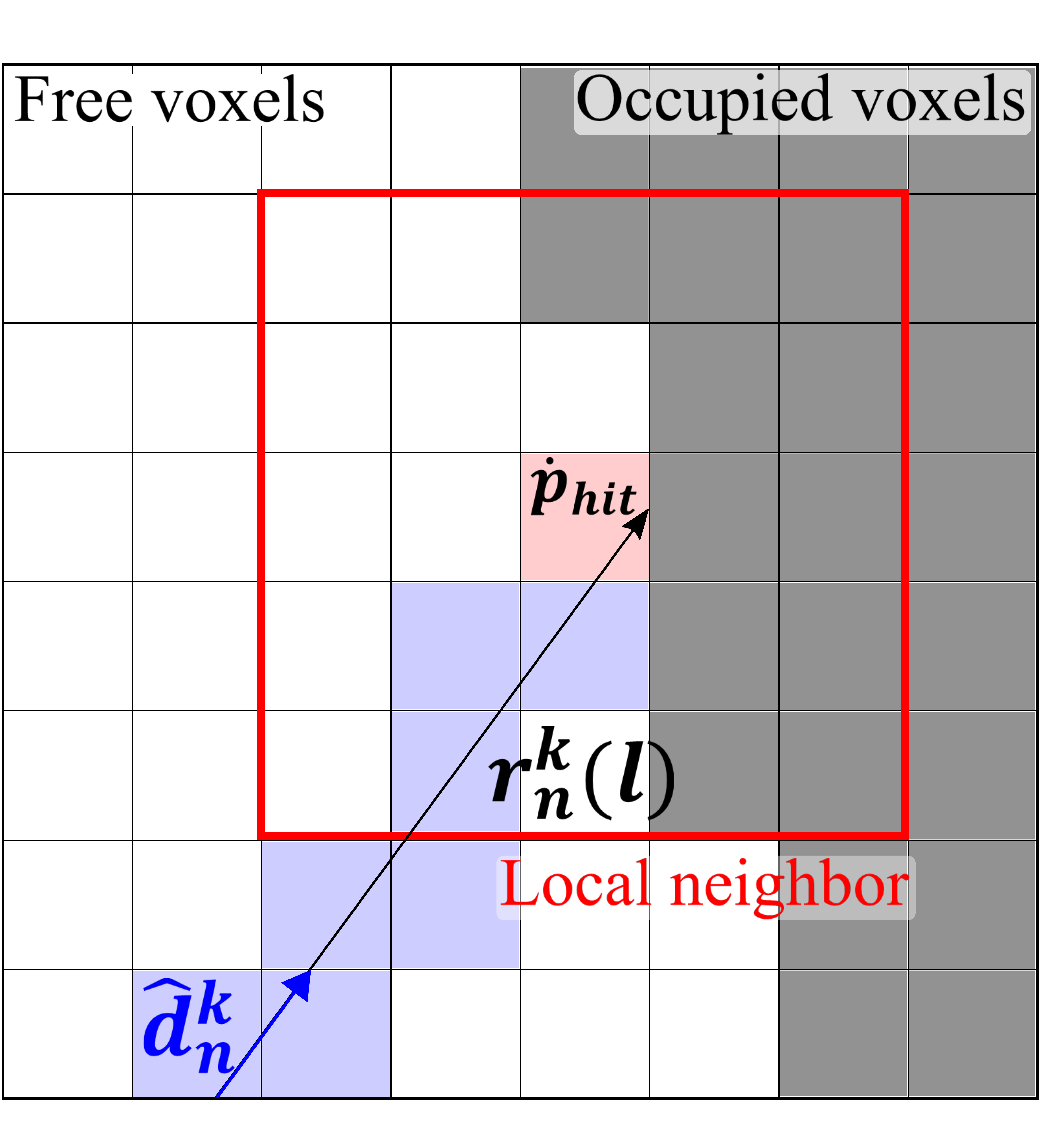}\label{fig:acousticRayTracing_hit}}\quad\,
	\subfloat[Computing a normal]{\includegraphics[width=1.52in]{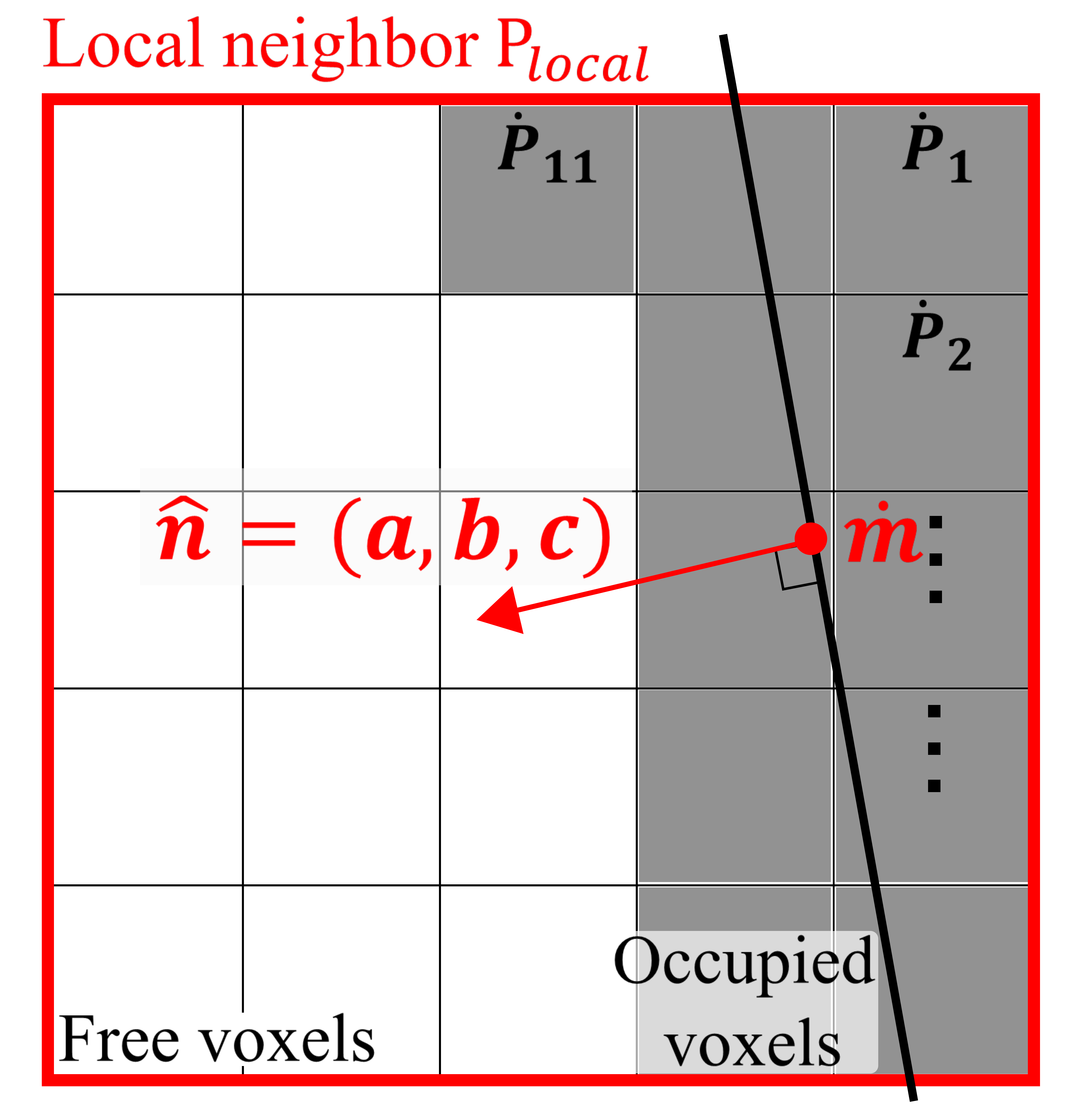}\label{fig:acousticRayTracing_compute_normal}}\quad\,
	\subfloat[Generating a reflection ray]{\includegraphics[width=1.5in]{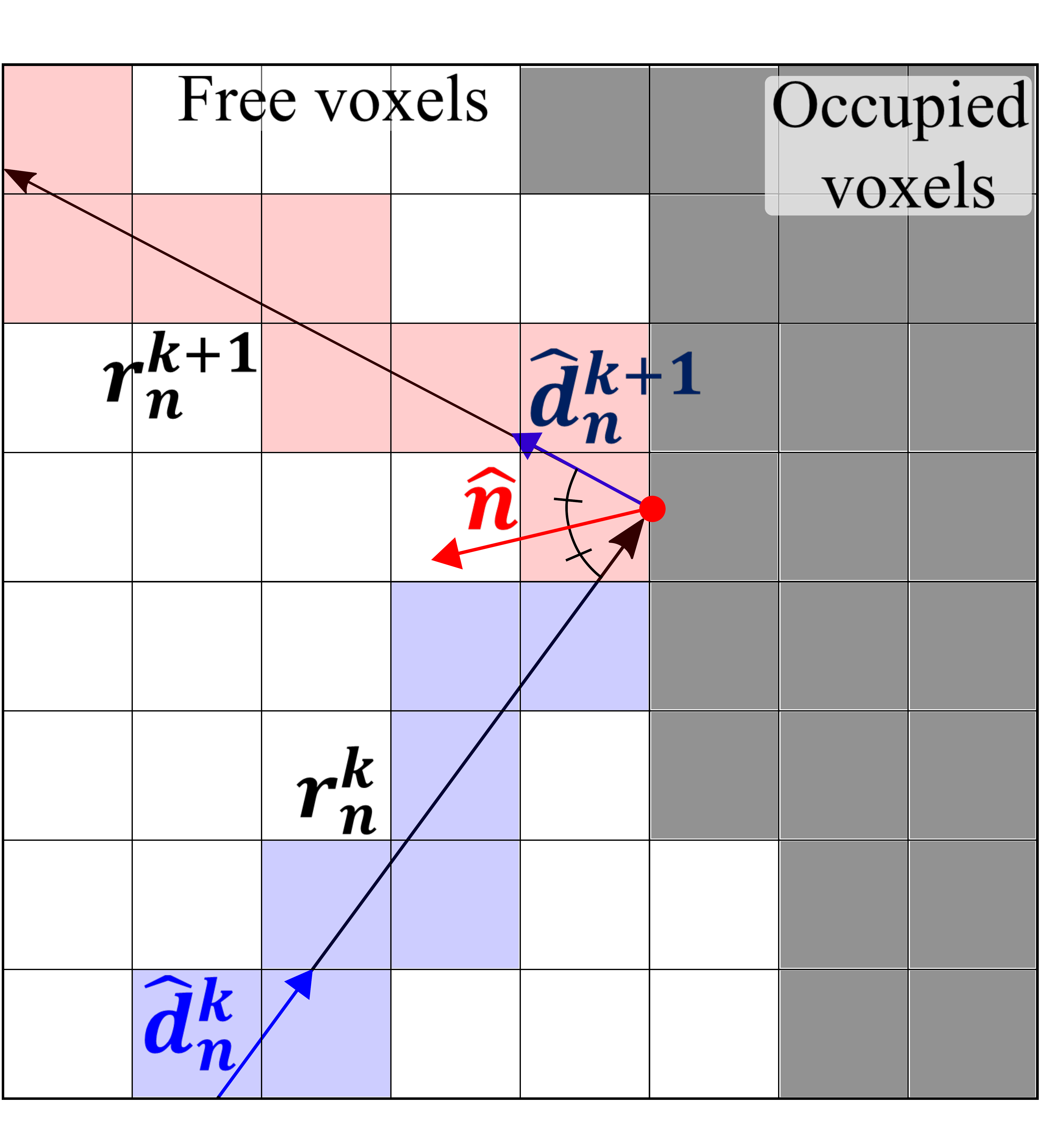}\label{fig:acousticRayTracing_reflect}}
	\Skip{
		\begin{subfigure}[b]{1.5in}
			\includegraphics[width=\textwidth]{figures/InvRT_1_Initialize}
			\vspace{0.1in}
			\vspace{-0.2in}
			\caption{Initializing an acoustic ray\label{fig:acousticRayTracing_init}}
		\end{subfigure}\quad
		\begin{subfigure}[b]{1.5in}
			\includegraphics[width=\textwidth]{figures/InvRT_2_HitEvent}
			\vspace{0.1in}
			\vspace{-0.2in}
			\caption{Detecting a hit\label{fig:acousticRayTracing_hit}}
		\end{subfigure}\quad
		\begin{subfigure}[b]{1.52in}
			\includegraphics[width=\textwidth]{figures/InvRT_3_NormalEstimation}
			\vspace{-0.2in}
			\caption{Computing a normal\label{fig:acousticRayTracing_compute_normal}}
		\end{subfigure}\quad
		\begin{subfigure}[b]{1.5in}
			\includegraphics[width=\textwidth]{figures/InvRT_4_Reflection}
			\vspace{0.1in}
			\vspace{-0.2in}
			\caption{Generating a reflection ray\label{fig:acousticRayTracing_reflect}}
		\end{subfigure}
	}
	
	\caption{ 
		This figure illustrates our acoustic ray tracing.  (a) An acoustic ray $r_n^{0}(l)$ is initialized inversely to an incoming sound direction. (b) Another acoustic ray $r_n^k(l)$, which is reflected $k$ times from its initial ray $r_n^0(l)$, is propagated and intersected with an obstacle encoded in the occupancy map. (c) On the fly, we compute a normal from a 2D plane, which locally fits the surface within its local neighbor cells, $P_{local}$, by using singular value decomposition. (d) From the hit point, we generate its reflected acoustic ray $r_n^{k+1}(l)$  in the direction of $\hat{d}_n^{k+1}$, assuming specular material at the hit point.
	}
	\vspace{-1em}
	\label{fig:acousticRayTracing}
\end{figure*}

To inversely determine how sound signals are received, we propose using acoustic ray tracing; technically, it is inverse acoustic ray tracing, but we choose just to call it acoustic ray tracing for simplicity. Note that the positions of a sound source and its listener can be interchanged thanks to the acoustic reciprocity theorem~\cite{brandstein2013microphone}. Fig.~\ref{fig:acousticRayTracing} shows the overview of our acoustic ray tracing, which is discussed in the following paragraphs.

\paragraph{Initialization.}
On each invocation of our method, we first run a TDOA module, which discretizes the captured sound signal into $N$ incoming sounds. An $n$-th incoming sound is represented by a tuple $(\hat{v}_n, f_n, i_n^0)$, where a unit vector $\hat{v}_n$ describes the incoming direction, $f_n$ indicates the representative frequency that has the highest energy of the incoming signal, and  $i_n^0$ represents its measured energy value of the sound pressure collected by the microphone array. We then generate an acoustic ray, $r_n^0$, by the following parametric equation with a ray length, $l \ge 0$: 
\begin{equation}
r_n^0(l) = \hat{d}_n^0 \cdot l + \dot o_m,
\label{eq:initial ray definition}
\end{equation}
where $\dot o_m$ represents the origin of the microphone array, and $\hat{d}_n^0$ is a directional unit vector in the inverted direction of the incoming sound, i.e., $\hat{d}_n^0 = -\hat{v}_n$. The superscript $k$ of an acoustic ray, $r_n^k(l)$, indicates the number or order of reflection along an acoustic path from the microphone array. For example, $r_n^0(l)$ indicates that there is no reflection and thus denotes a direct ray from the microphone array. All the other rays with a varying number of reflections, i.e. $k \ge 1$, are called indirect acoustic rays with $k$-th order reflections.

\paragraph{Propagation in the empty space.}
Once an acoustic ray is generated, it is propagated through space and can be
reflected once it hits an obstacle. During this acoustic ray tracing process,
we have to amplify the energy of the acoustic ray to simulate the
propagation and reflection operations.

In particular, an energy function, $I_n^k (l')$, of a ray $r_n^k$ at a particular ray length, $l'$,  i.e. $l=l^\prime\enskip(l^\prime \geq 0)$, is defined as follows:
\begin{equation}
I_n^k( l^\prime ) = i_n^k \cdot \exp(\alpha(f_n) l^\prime),
\label{eq:energy propagation}
\end{equation}
where $i_n^k$ is the initial acoustic energy of the ray at $l^{\prime}=0$, and $\alpha (f_n)$ is the attenuation coefficient, which depends on the frequency of the sound $f_n$, and other environment-related factors such as temperature and humidity of the air. Our formulation is based on an inverse operation of the normal decay of the sound signal~\shortcite{siltanen2007room}.

\paragraph{Specular reflection.}
When a ray $r_n^k$ hits the surface of an object in the scene,
we need to simulate how the ray behaves at the hit point.
Ideally, reflection, absorption, or diffraction occurs, depending on the
material type of the hitting surface. Since simulating all these types of
interactions requires a prohibitive computation time, we only support on
absorption and reflection in this work assuming high-frequency sound signals,
e.g., higher than 2~kHz.  In terms of reflection, there commonly exist specular
and diffuse acoustic materials.  We also assume the specular material type and
generate our reflected acoustic rays based on that material.

Our choice to not support diffuse reflections is based on two factors: 1) supporting diffuse reflections requires an expensive inverse simulation approach such as Monte Carlo simulation, which is unsuitable for real-time robotic applications, and 2) while there are many diffuse materials in rooms, each individual sound signal reflected from the diffuse material does not carry a high portion of the sound energy generated from the sound source.   Therefore, when we choose high-energy directional data from the TDOA based method, the most sound signals reflected by the diffuse material are ignored automatically, and those with high energy are mostly from specular materials.

Note that our work does not require all the materials to be specular. When some of the materials exhibit high energy reflectance near the specular direction, e.g., tex materials in the ceiling and finished wooden floors, our method generates acoustic rays toward those directions, and our detection method will identify the location of the sound source that generates those rays.  As a result, we focus on handling specular materials well and treat each hit material as specular, and generate a reflected ray from the hit point.



\Skip{
	When the material type is not exactly specular,
	reflected rays from it may not be useful for localizing the sound source
	assuming that there are some specular materials
	In reality, when rooms have near specular materials such as\YOON{refine examples}\IK{Addressed}, our
	generated rays can account for the meaningful energy reflections from those
	materials.
}


The operation for specular reflection is defined as follows. Whenever a previous acoustic ray, $r_n^{k}$, hits the surface of the obstacle at the particular ray length, $l_{hit}$, we create a new, reflected acoustic ray, $r_n^{k+1}$, with the following direction and energy equations:
\begin{equation}
\begin{aligned}
r_n^{k+1}(l) &= \hat{d}_n^{k+1} \cdot l + r_n^k(l_{hit}), \\
i_n^{k+1}\enskip &= I_n^k ( l_{hit} )\, /\, ( 1 - \alpha_s ),
\end{aligned}
\label{eq:reflection}
\end{equation}
where $\hat{d}_n^{k+1}$ is the direction of the specular direction of the ray 
$r_n^{k+1}$, and 
is analytically computed by $\hat{d}_n^{k+1} = \hat{d}_n^k - 2(\hat{d}_n^k \cdot \hat{n})\hat{n}$, 
where $\hat{n}$ is the normal vector at the surface hit point $r_n^{k+1}(0)$.
Also, $i_n^{k+1}$ is its initial energy. 
The absorption coefficient, $\alpha_s$, describes the energy lost on the
surface during the reflection~\shortcite{schissler2017acoustic}.

The reflection ray that we create can be reflected further by getting another hit on other obstacles.  This recursive reflection process is terminated when the energy of a ray, $i_{n}^{k}$, exceeds a user-defined threshold for maximum energy, denoted as $i_{thr}$, which is set by a reasonable energy bound, i.e., 900~J that we can hear in most indoor scenes. While generating the acoustic rays of a path, we maintain them in a ray sequence, $R_n = [ r_n^0, r_n^1, ... ]$ generated for the $n$-th incoming sound.  We use this ray sequence to estimate the location of the sound source.

\begin{figure}[tp]
	\centering
	\subfloat[]{\includegraphics[width=0.45\columnwidth]{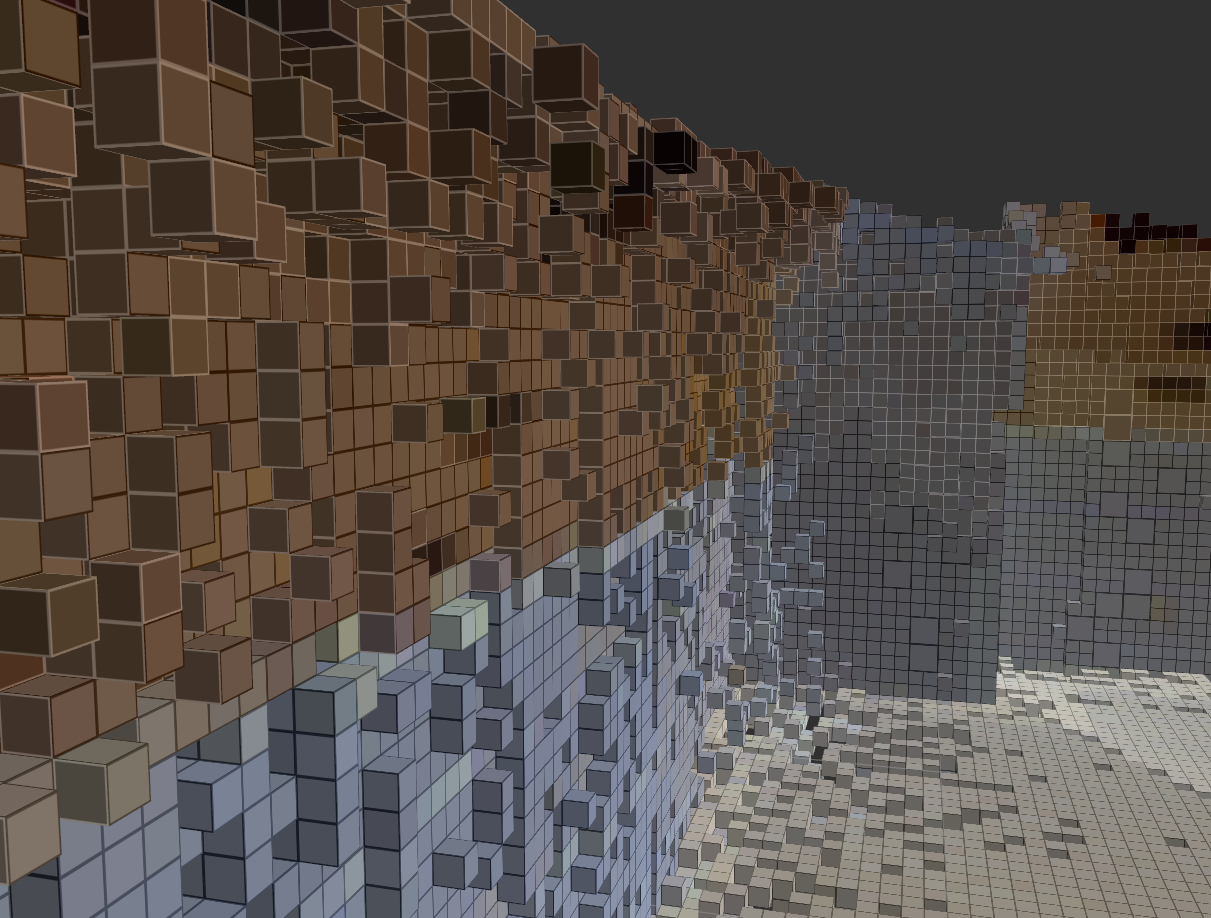}\label{fig:smoothing_before}}\qquad
	\subfloat[]{\includegraphics[width=0.45\columnwidth]{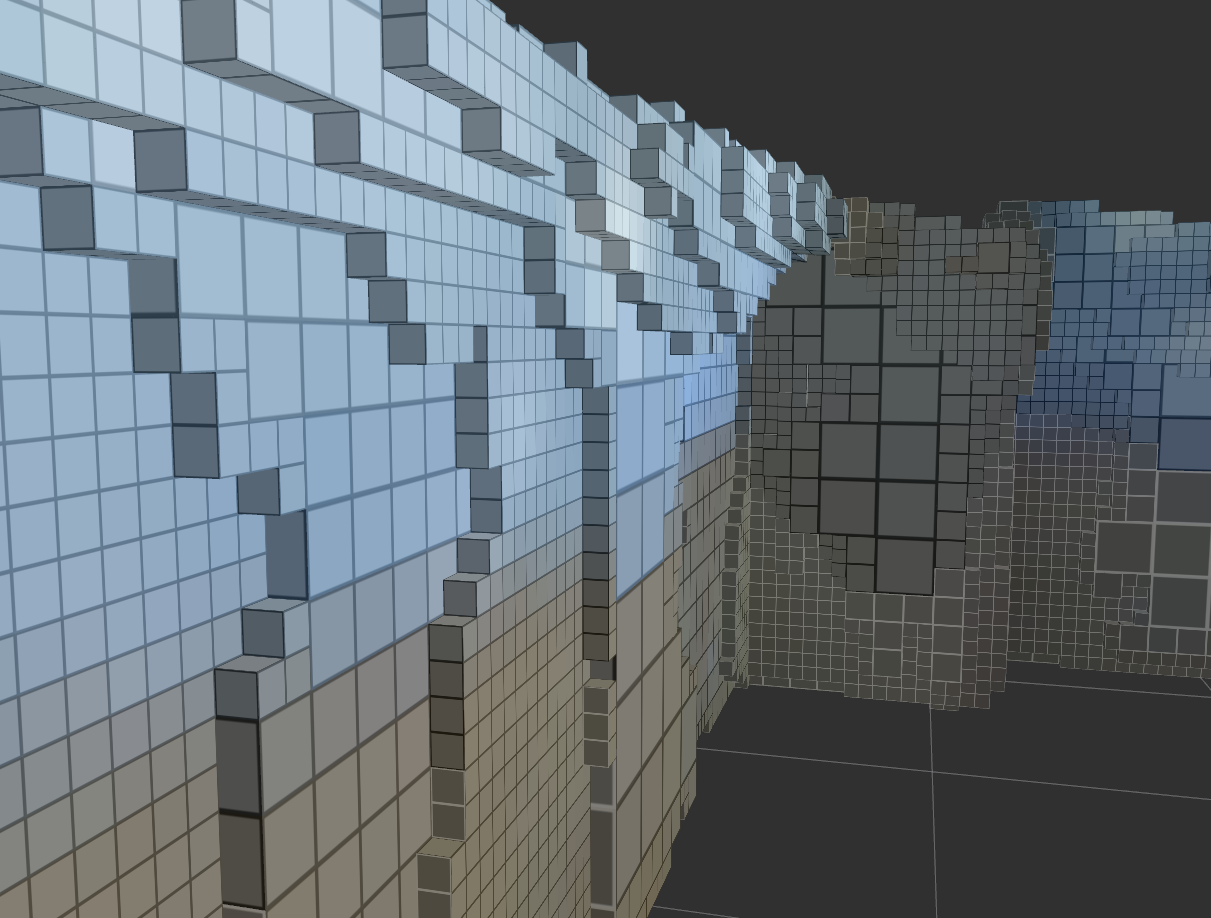}\label{fig:smoothing_after}}
	\caption{
		(a) and (b) show the original voxels of the wall that have a
		high level of noise and the voxels refined by our SVD based
		approach, respectively.
	}
	\vspace{-1em}
	\label{fig:smoothing}
\end{figure}

\paragraph{Smoothing octree map.} 
As in other practical robotics applications, we use the octree map representation for the reconstructed 3D space, and perform our acoustic ray tracing with it.  Unfortunately, the underlying map structure may contain a high level of noise even though we use high-quality sensors.  Such noises can make rough surfaces and thus varying normals of the surfaces, resulting in low quality in terms of tracking acoustic paths and identifying the sound source (Fig.~\ref{fig:smoothing_before}).

To address this issue, we propose using a simple, yet effective low-pass filter using singular value decomposition (SVD) that works in an on-the-fly manner. Given a cell $\dot{p}_{hit}$ intersected by an acoustic ray, we identify  a set of local neighbor voxels, $P_{local}$, which include occupied cells in a cubic volume  centered at the cell $\dot{p}_{hit}$ (Fig.~\ref{fig:acousticRayTracing_compute_normal}). We then compute $\dot{m}$, the average position of those occupied voxels of $P_{local}$, and a matrix $A$, each column of which contains a vector from $\dot{p}$ to the center of each occupied voxel. Our goal is then to compute a vector $\hat{n}_s$ among possible normal vectors $\hat{n}$ that minimizes the Euclidean norm  of vector angles between the normal vector and vectors in the matrix A, which is formulated as the following:

\begin{equation}
\begin{aligned}
\hat{n}_s&= \argmin_{\hat{n}} \| A^T\hat{n} \|_2
= \argmin_{\hat{n}} \| V S^T U^T \hat{n} \|_2 \\
&= \argmin_{\hat{n}} \| S^T U^T \hat{n} \|_2 = U^T(3,:),
\label{min_formula_SVD}
\end{aligned}
\end{equation}
where $V S^T U^T$ is computed by SVD~\cite{golub1970singular}.
It is well known that $\| S^T U^T \hat{n} \|_2$ has the maximum value when
$\hat{n}$ equals $U^T(3,:)$, the eigenvector with the smallest eigenvalue.

Fig.~\ref{fig:smoothing} shows that our simple on-the-fly smoothing process shows significantly improved quality over the one without the smoothing operation.  Overall, our SVD based computation runs quite fast and takes only 0.07\% of the overall computation. Note that reconstructing a high-quality representation itself is one of the active research areas and ours can be improved by alternatives, e.g., extracting a high-quality surface.

\subsection{Identifying a Converging 3D Point}
\label{sec:4_b}

So far we generated direct and reflected acoustic rays starting from incoming sound signals.  Given those acoustic ray paths, we are ready to localize a sound source in the 3D space. For the sake of clarity, we assume that all sound signals originate from a single sound source; handling multiple targets using a particle filter has been well studied~\cite{okuma2004boosted}, and can be used for our approach.

In an ideal case, it is sufficient to find a point at which acoustic rays intersect. However, since we deal with real environments in practice, there are diverse types of noise from sensors (e.g., microphones and Kinect), and we need a technique that is robust to those types of noise. As a result, we cast our problem as locating a region where many of those ray paths converge. Once the region is small enough, we treat the region as containing the sound source.

For achieving our goal, we propose using  {Monte Carlo localization} (MCL)~\cite{thrun2005probabilistic}, also known as the particle filter, for localizing and representing such a region with particles.  Our localization method  consists of three parts: {sampling}, {weight computation}, and {resampling}. 

\paragraph{Sampling.}
Sampling starts with $N$ acoustic ray paths, $\{R_1, \dots, R_N\}$, generated by our acoustic ray tracing. At each sampling iteration step $t$,  we maintain a set of $W$ particles, $\chi_t = \{ x_t^{1} ,\cdots, x_t^{W} \}$, which serve as hypothetical locations of a sound source and are spread out randomly at the initial step in the 3D space. We associate a weight with each particle, and the weight is set to indicate its importance, specifically encoding how closely the particle is located to a nearby acoustic ray; we aim to re-generate more particles closer to those rays to achieve a higher accuracy in localizing the sound source.

For each iteration $t$ other than the initial iteration,
a new set of particles, $\chi_{t+1}$, is incrementally created
from the prior particles.  Specifically, a new particle, $x_{t+1}^i$, is generated
by offsetting an old one, $x_{t}^i$, in a random unit direction,
$\hat{u}$, as an offset, $d$, as in the following:
\begin{equation}
x_{t+1}^{i} = x_{t}^{i} + d \cdot \hat{u},
\label{prediction_PF}
\end{equation}
\begin{equation}
d = \| x_{t+1}^{i} - x_{t}^{i} \| \thicksim {N}(0, \sigma_s
),
\end{equation}

where $N(\cdot)$ denotes a normal distribution, the mean of which is zero and the std. deviation of which is determined by the size of the environment; 1~m is set to $\sigma_s$ for 7~m by 7~m room space.

\begin{figure}[t]
	\centering
	\includegraphics[width=1.7in]{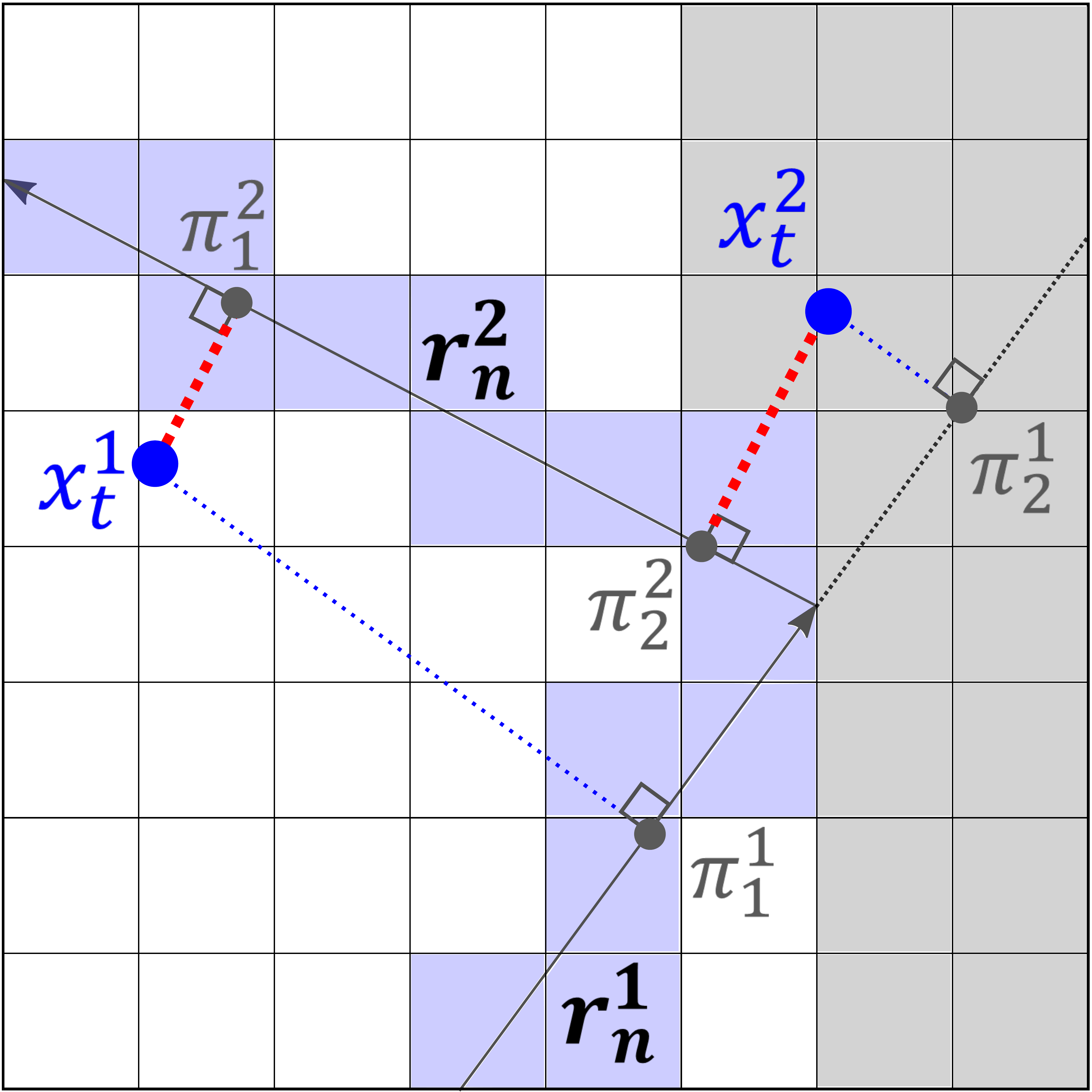}
	\caption{
		This figure shows an example of computing weights for
		particles against a ray path, $R_n=[r_n^1, r_n^2]$.  The chosen
		representative weight for each particle is shown in the red
		color.
	}
	\vspace{-1em}
	\label{fig:compute_distance_particle}
\end{figure}


\begin{figure*}[t]
	\centering
	\subfloat[Our robot.]{\includegraphics[width=0.26\columnwidth,valign=c]{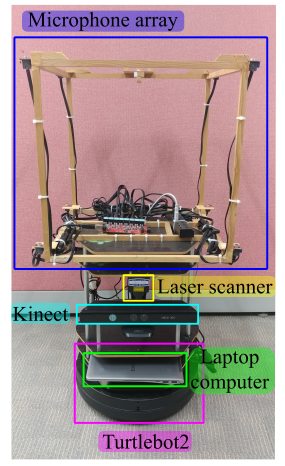}\label{fig:hardware}}\,
	\subfloat[Stationary sound source.]{\includegraphics[width=0.80\columnwidth,valign=c]{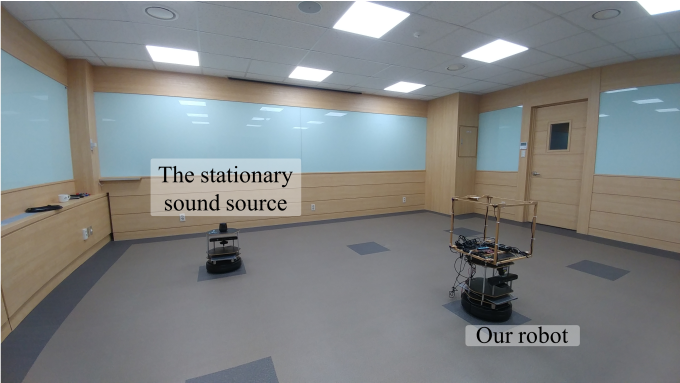}\label{fig:staticSoundEnv}}\,
	\subfloat[Moving sound w/ an
	obstacle blocking the line-of-sight.]{\includegraphics[width=0.85\columnwidth,valign=c]{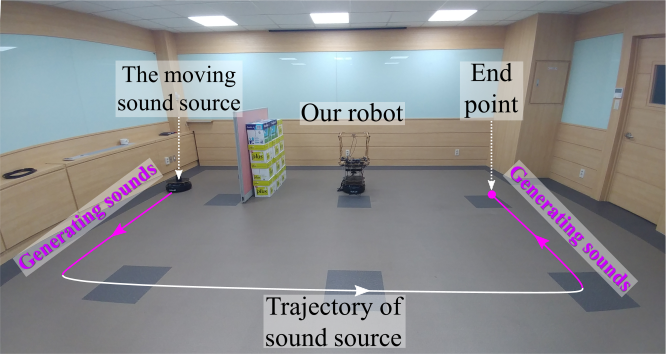}\label{fig:DynamicSoundEnv}}
	\Skip{
		\begin{subfigure}[b]{\columnwidth}
			\includegraphics[width=\textwidth]{figures/Experiment_StaticSound_Picture}
			\vspace{-1em}
			\caption{Static sound source\label{fig:staticSoundEnv}}
		\end{subfigure}\\
		\begin{subfigure}[b]{\columnwidth}
			\includegraphics[width=\textwidth]{figures/Experiment_DynamicSound_Picture}
			\vspace{-1em}
			\caption{Dynamic robot w/ obstacle blocking the line-of-sight.\label{fig:DynamicSoundEnv}}
		\end{subfigure}
	}
	\caption{ 
		(a) shows our tested robot with the cube-shaped microphone array.
		(b) and (c) show our testing environments for static and
		dynamically moving sound sources, respectively.
		For the moving sound, it generates sounds, only when it is on
		the violet part of its trajectory.
	}
	\vspace{-1em}
	\label{fig:Environment}
\end{figure*}

\paragraph{Weight computation.}
In this step, we compute the likelihood of the $i$-th particle given the acoustic rays.  Since we want to generate particles close to acoustic rays, we assign a higher weight to a particle when the particle is more closely located to the rays.  Specifically, given the observation of ray paths, $\mathbf{o}_t = [ R_1, R_2, \cdots, R_N ]$, we define the likelihood $P( \mathbf{o}_t|x_t^i)$ as follows:
\begin{equation}
P(\mathbf{o}_t| x_t^i) = \frac{1}{n_c} \sum_{n=1}^N \left\{\max_k {w(x_t^{i}, r_n^k)}\right\}, \\
\label{important_weight_likelihood_PF}
\end{equation}
where a weight function, $w$, is defined between
a particle $x_t^{i}$ and a ray $r_n^k$, the $k$-th order reflection ray of
the $n$-th ray path $R_n$, and $1/n_c$ is a normalization factor over the
likelihood of all particles.  Simply speaking, for each  particle, we pick a
representative weight as the
maximum weight among weights computed from rays  in each ray path and
accumulate the representative weights with all the ray paths.
In the example shown in Fig.~\ref{fig:compute_distance_particle}, there are two
rays, $r_n^1$ and $r_n^2$, with an acoustic path $R_n$. If a particle $x_t^{1}$
is closer to $r_n^2$ than $r_n^1$ on their acoustic path $R_n$, $w(x_t^{1},\,
r_n^2)$ is chosen as the representative weight contribution
for the ray path $R_n$.

The weight function $w(x_t^{i}, r_n^k)$ is defined as follows:
\begin{equation}
\begin{aligned}
w(x_t^{i}, r_n^k) &= {f_N}(\|x_t^{i} - \pi_i^k \|\ |\ 0, \sigma_w) \times {F}(x_t^{i}, r_n^k),
\end{aligned}
\end{equation}
where $\pi(x_t^i,r_n^k)$, in short, $\pi_i^k$, returns the perpendicular foot of the particle $x_t^{i}$ to the ray $r_n^k$ (Fig.~\ref{fig:compute_distance_particle}), and $f_N(\cdot)$ denotes the pdf of the normal distribution. $\sigma_w$ is set according to the determinant of the covariance matrix of particles; as a result, we assign a higher weight to a particle close to a ray, as other particles are more distributed. ${F}$ is a filter function returning zero to exclude irrelevant cases when the perpendicular foot is outside of the ray segment $r_n^k$, e.g., $\pi_2^1$ in Fig.~\ref{fig:compute_distance_particle}; otherwise, the filter function returns one.
%

\paragraph{Resampling.}
The likelihood weight associated with each sampled particle $P(\mathbf{o}_t | x_t^{i})$ is used to compute an updated set of particles for the next step $t+1$.  Intuitively, in this process, particles with low weights are removed, and additional particles are generated near the existing particles with high likelihood weights. For this process, we adopt a basic resampling method~\cite{thrun2005probabilistic}. 

Once resampling is done, we check whether particles are converged enough to define an estimated sound source.   To determine the convergence of the positions of particles, we compute the generalized variance (GV), which is a one-dimensional measure for multi-dimensional scatter data and is defined as the determinant of the covariance matrix of particles~\cite{anderson1984introduction}. If GV is less than the convergence threshold, $\sigma_c=0.01$, we terminate our process and treat the mean position of the particles as the estimated position of the sound source. GV is also used as a confidence measure on our estimation; we also use its covariance matrix to draw 95\% confidence ellipsis disk for visualizing the estimated sound region (Fig.~\ref{fig:resultOfAcousticRay}).

\section{RESULTS and DISCUSSIONS}
\label{sec:5}

In this section, we explain our tested robot with a microphone array and
environments, followed by demonstrating the benefits of our method. 

\paragraph{Hardware setup.}
Fig.~\ref{fig:hardware} shows our tested robot used for localizing the sound source. This robot is based on a Turtlebot2 equipped with three types of sensors: Kinect, Laser scanner, and microphone array.  Kinect and Laser scanner generate RGB-D and point cloud streams passed down to the SLAM module, RTAB-Map method \shortcite{labbe2014online}, as shown in Fig.~\ref{fig:blockDiagram}.  The resulting environment is represented in Octomap~\shortcite{hornung2013octomap}, an octree-based occupancy representation.

The robot receives the sound stream from the microphone array, which is an embedded auditory system introduced in \shortcite{briere2008embedded}, and generates directions of sound signals based on a TDOA-based method utilizing ManyEars open software~\shortcite{grondin2013manyears}. We use a clapping sound as the sound source that has frequencies higher than 2kHz.  All of our methods are processed in the laptop computer built in the robot, which includes an Intel i7 processor 7500U with 8GB memory.

\paragraph{Testing scenarios.}
To demonstrate the benefits of our reflection-aware method, we test our approach in three different testing scenarios in a classroom environment (Fig.~\ref{fig:Environment}): 1) a stationary sound source with continuous sound signals,  2) a stationary sound source with intermittent sound signals,  and 3) a moving sound source with intermittent signals.   Most prior approaches focused on finding a sound source with accumulated sound data, while their robots are moving~\cite{narang2014auditory,sasaki2016probabilistic,su2017towards}.   Along with this prior benchmark setting, we include the first scenario, in which we can accumulate the sound data with the continuous signal. Note that many types of sounds in the real world are frequently generated in an intermittent manner rather than in a continuous manner; e.g., a human can call robots by voice or clapping, which can be classified as intermittent signals. As a result, we include the second and third benchmarks, where sound signals are intermittent. These two scenarios are challenging cases that were not tested in most prior approaches.  Furthermore, many prior approaches do not consider the moving sound source of the third benchmark, which hinders the accumulation of sound signals~\cite{narang2014auditory,sasaki2016probabilistic,su2017towards}. Because our method efficiently considers reflection, it can handle such challenging cases.

\subsection{Environments with a moving robot and an obstacle}
\label{sec:5_a}

We first show results with a stationary sound source generating continuous or intermittent sound signals (Fig.~\ref{fig:Environment}(b)).   At each running of our method, we generate 60 acoustic rays on average, and we show only top-3 acoustic ray paths regarding its carried energy for the clear visualization in Fig.~\ref{fig:resultOfAcousticRay}. We can see a strongly reflected ray from the ceiling, with other directed rays from the source. Thanks to these strong direct and reflected rays passing through the region, our particle filter can detect the location of the sound source well. Note that there are also acoustic ray paths that do not pass the identified region, but their intensities are small, i.e., about 50\% compared to the average of those top-3 ray paths.

\begin{figure}[t]
	\centering
	\subfloat[Continuous sound]{\includegraphics[width=0.9\columnwidth]{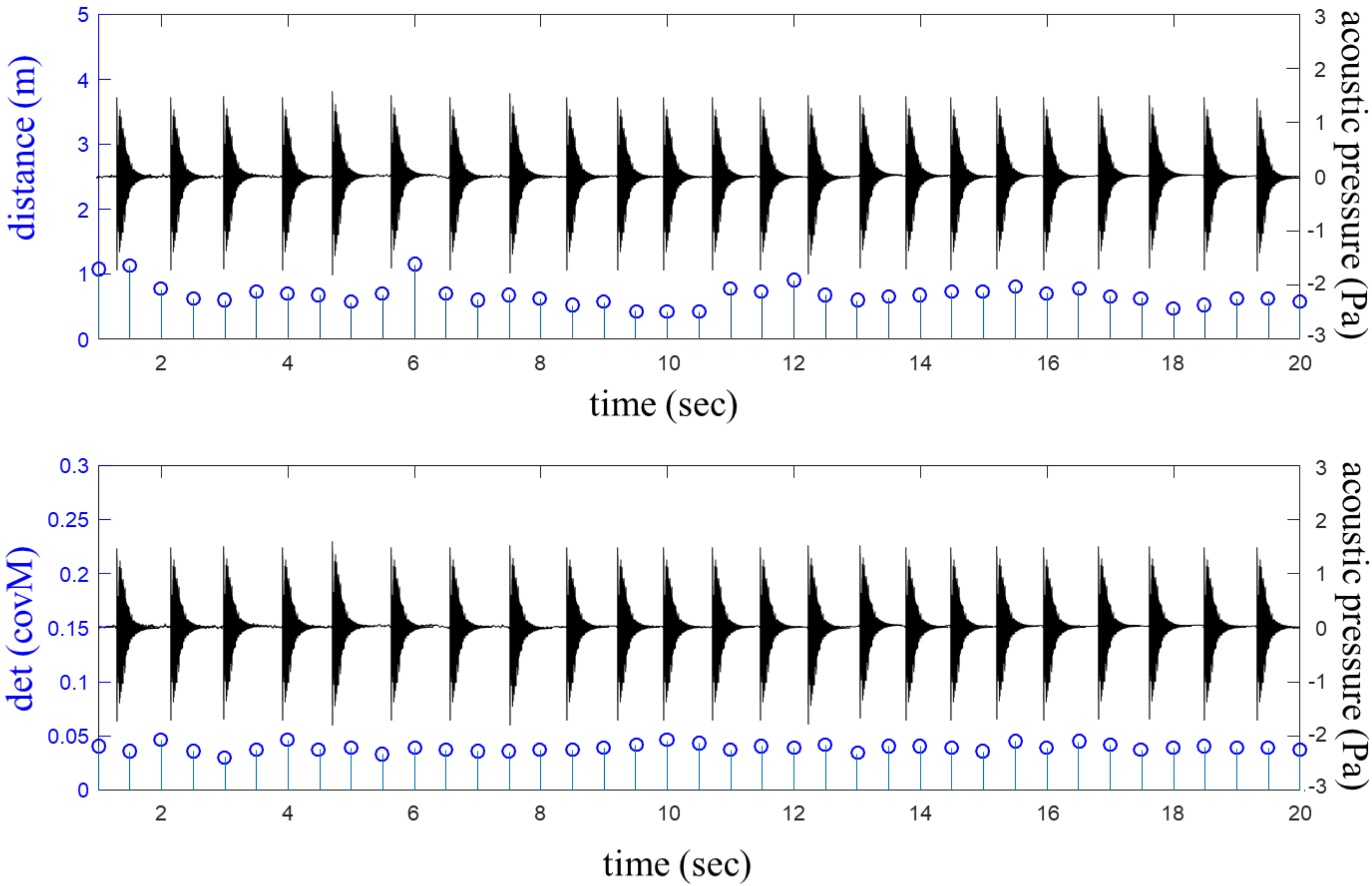}\label{fig:continueResult}}\\
	\subfloat[Intermittent sound]{\includegraphics[width=0.9\columnwidth]{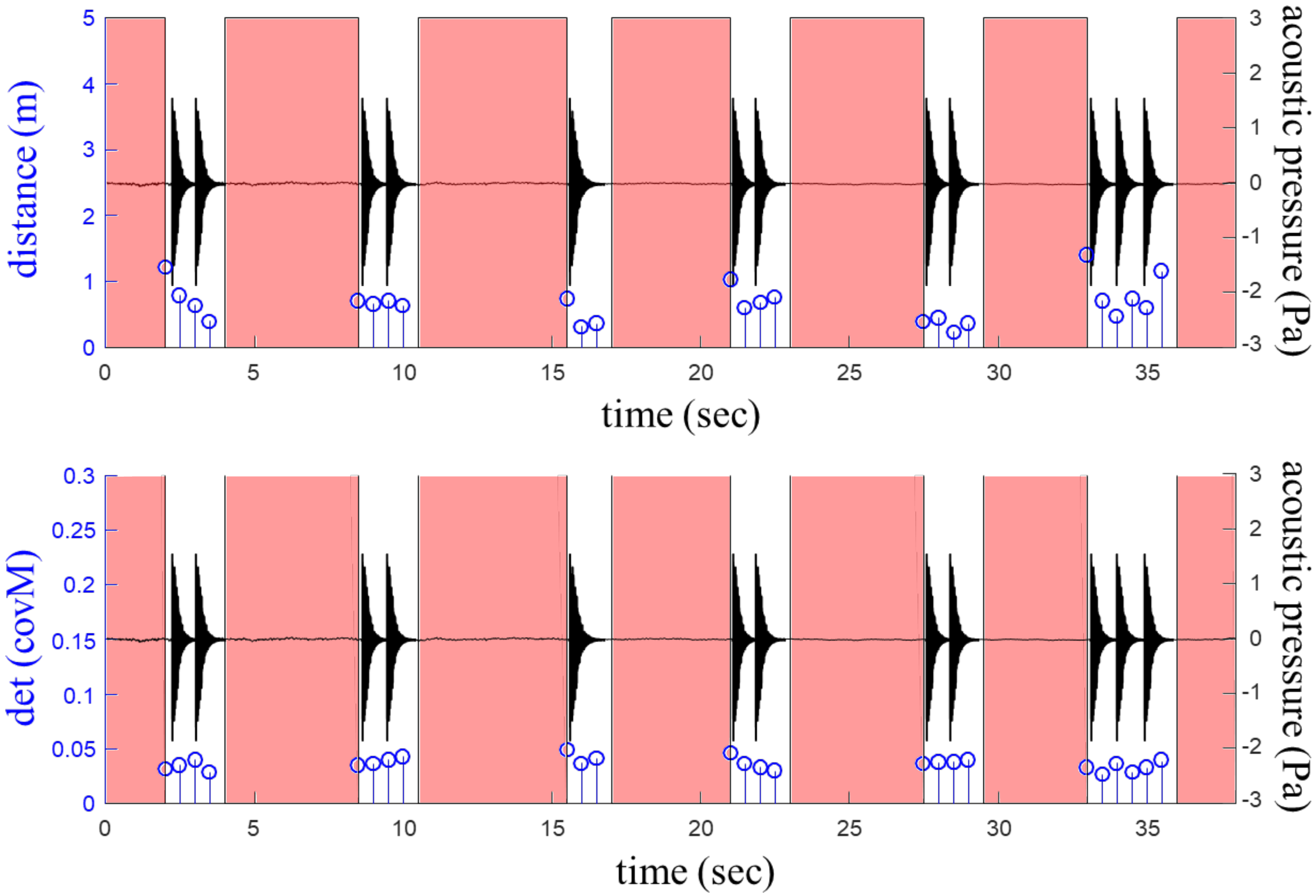}\label{fig:intermittentResult}}
	\Skip{
	\begin{subfigure}[b]{\columnwidth}
		\includegraphics[width=\textwidth]{figures/Experiment_StaticSound_Continuous_ErrorGraph}
		\vspace{-1em}
		\caption{Continuous sound\label{fig:continueResult}}
	\end{subfigure}\\
	\begin{subfigure}[b]{\columnwidth}
		\includegraphics[width=\textwidth]{figures/Experiment_StaticSound_Intermittent_ErrorGraph}
		\vspace{-1em}
		\caption{Intermittent sound\label{fig:intermittentResult}}
	\end{subfigure}
	}
	\caption{ 
		This graph shows the results of the average error distance and the determinant of the covariance matrix with the stationary sound source. The avg. error distance is measured between the ground truth and the estimated position in the 3D space. For the intermittent case, (b), the red background is used when we do not have any signals. Acoustic pressure of the measured sound signals is also shown.
	}
	\vspace{-1em}
	\label{fig:resultGraph}
\end{figure}

Fig.~\ref{fig:resultGraph} shows the average distance error between the ground truth and the estimated sound location, with determinants of the particle covariance matrix. For the continuous case, the mean and standard deviation of the distance errors are 0.72~m and 0.26~m, respectively. The standard deviation is quite small, indicating that our method stably determines the sound location from generated acoustic rays. The average error 0.72~m is slightly large compared to its std. value. This error is mainly attributed to bias, which is caused by various factors such as reconstruction errors of SLAM, the TDOA-based method, and errors of our method, which does not consider characteristics of low frequencies of sound signals. Nonetheless, the average error of  72~cm is reasonably useful for our robotics application. Also, the determinants of the covariance matrix for the particle filter are very small (less than 0.1), indicating that the particles in the particle filter are converged well.

For the intermittent case, we toggle the sound generation in every 5 seconds. The mean and std. deviation of the distance errors are 0.66~m and 0.29~m, respectively. This result is similar to the continuous case, and it shows that our algorithm localizes the intermittent sound source well, when the sound source is stationary.  The determinants, in this case, are also small.

\subsection{Environments with a dynamic sound source and obstacles}
\label{sec:5_c}

Fig.~\ref{fig:DynamicSoundEnv} shows the trajectory of the moving sound source. To make it a more challenging benchmark, we also put an obstacle on the left side of the robot, and the sound occurs only when the sound source is in the violet part of the trajectory.

\begin{figure}[t]
	\centering
	\subfloat[Detected regions as the sound source moves.]{\includegraphics[width=0.8\columnwidth]{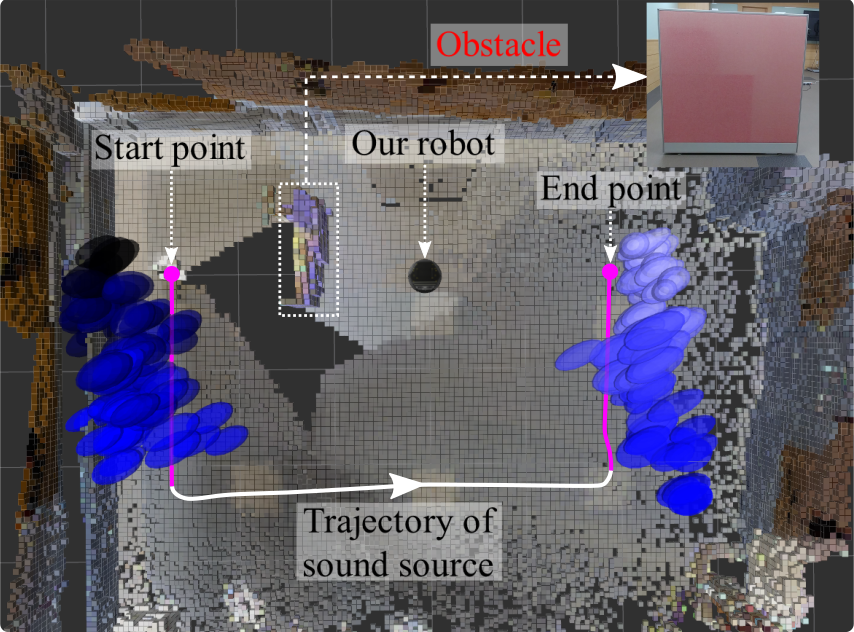}\label{fig:Experiment_DynamicSound_Visualization}}\\
	\subfloat[Measured distance error.]
	{\includegraphics[width=0.9\columnwidth]{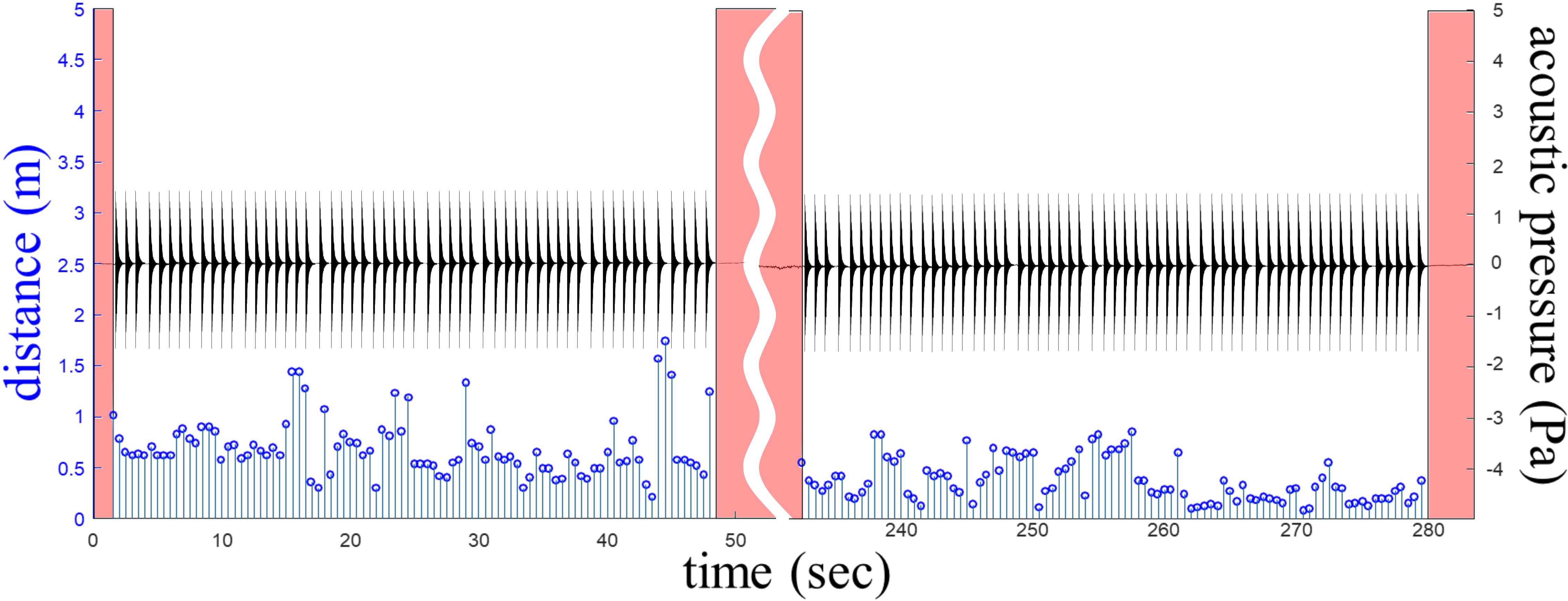}\label{fig:Experiment_DynamicSound_ErrorGraph}}
	\caption{
		(a) shows detected regions as the sound source moves in the environment of Fig.~\ref{fig:DynamicSoundEnv}; we change the color of the detected disk from the dark blue to light one as the time passes. Note that the source does not generate any sound, while it is in the lower middle part of the trajectory. (b) shows the distance error as a function of the time on the trajectory; we use the red background when we do have any sound.
	}
	\vspace{-1em}
	\label{fig:DynamicSoundSource}
\end{figure}

Fig.~\ref{fig:DynamicSoundSource} shows the detected regions of the sound source as it moves. The lower graph of Fig.~\ref{fig:DynamicSoundSource} shows the distance error as a function of time.  The distance errors from 1 s to 50 s are measured when the source is located on the left side of the robot, while the errors from 230 s to 280 s are from the right side. The average error, 0.7~m, on the left side is higher than that,  0.3~m, of the right side.  The lower error on the left side is caused since the obstacle on the left side causes diffraction and reverberation, which decrease the detection accuracy of our method. Nonetheless, our method can generate reflected rays towards the sound source, while direct paths from the source are blocked due to the obstacle.  As a result, its error even in the very challenging case with the obstacle and moving sound is within a reasonable bound.  Furthermore, the std. deviations of the left (0.29m) and right (0.20m) sides are reasonably small, indicating that our method can stably identify the location of the sound source. 

\begin{figure}[t]
	\centering
	\includegraphics[width=1.0\columnwidth]{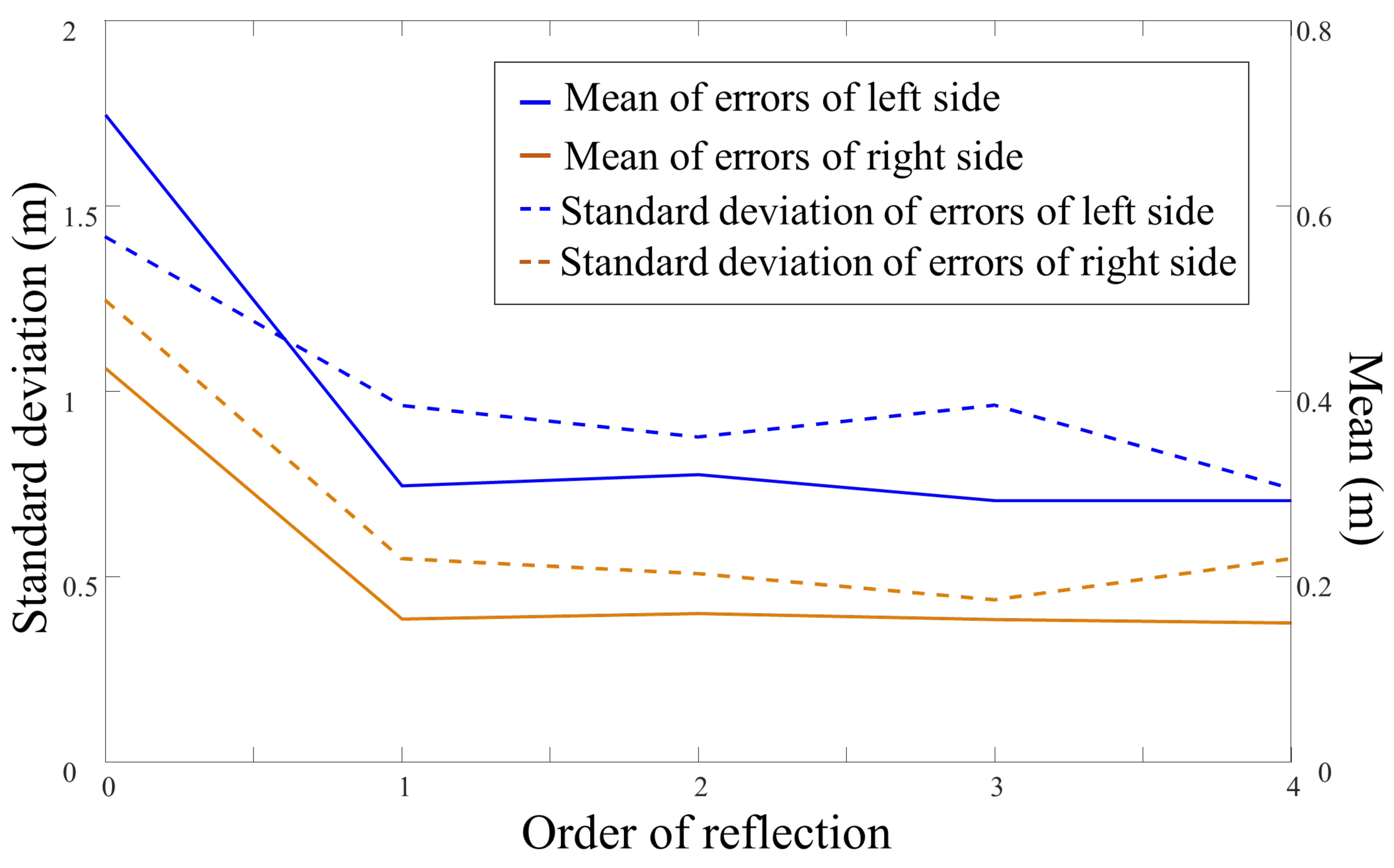}
	\caption{
		This graph shows the average distance error and its std. deviation as a function of the accumulated orders of reflection in the third benchmark with the obstacle; i.e., 1st reflection includes 1st reflection with the direct path. The result for left and right sides of the trajectory is separated.
	}
	\vspace{-1em}
	\label{fig:graphOfNumberOfReflection}
\end{figure}

\paragraph{Accuracy with the reflection order.}
To see the benefits of considering reflected rays in addition to direct rays, we measure the accuracy as a function of the accumulated orders of reflection rays. Fig.~\ref{fig:graphOfNumberOfReflection} shows the average distance error and std. deviation for the third benchmark with the moving source.  Especially, we measure such quantities separately for the left and right sides, to see their different characteristics.   The result of the right side is always better than that of the left side because the obstacle is closely located on the left side. When we consider the 1st order reflection additionally from the direct rays, various results are significantly improved, clearly demonstrating the benefit of considering reflected acoustic rays.  As we consider higher orders, we can also observe small, but meaningful improvements, in particular for the left side. Based on this result, we set the maximum order of reflections to be four in all of our tested benchmarks.  

\section{CONCLUSIONS \& FUTURE WORK}
\label{sec:6}

We have presented a novel, reflection-aware sound source localization algorithm based on acoustic ray tracing and Monte Carlo localization.  Thanks to the efficiency and considering direct and reflected acoustic paths, our algorithm can work with a single input frame without the accumulation of sound signals and can handle a moving sound source with an obstacle occluding the line-of-sight between the listener and sound source. We have evaluated these characteristics in a room with different source characteristics and configurations.  Furthermore, the use of reflected rays increases the localization accuracy substantially. 

While our results are promising, our approach has some limitations. It is mainly designed for high-frequency sources and does not model low-frequency effects like diffraction. Furthermore, our ray tracing model only takes into account specular reflections. As part of future work, we plan to accommodate wave-based approaches to improve the accuracy.  Another key issue is to have an accurate 3D reconstruction of the scene and to classify acoustic materials that affect the reflections. Finally, we would like to extend them to multi-source localization.

\section*{Acknowledgment}

{
\bibliographystyle{ieee/ieee}
\bibliography{Bib/robotics,Bib/soundSourceLocalization,Bib/Statistics,Bib/raytracing}

}

\end{document}